

\font\titolino=cmbx10
\font\tsnorm=cmr10
\font\tscors=cmti10

\font\tscorsp=cmti9
\magnification=1200

\hsize=148truemm
\hoffset=10truemm
\parskip 3truemm plus 1truemm minus 1truemm
\parindent 8truemm
\newcount\notenumber
\def \S{Schr\"o\-din\-ger }

\def\PR{{\tscors Phys. Rev. }}
\def\PRD{{\tscors Phys. Rev. D }}

\def\PLA{{\tscors Phys. Lett. A }}
\def\PLB{{\tscors Phys. Lett. B }}

\def\JMP{{\tscors J. Math. Phys. }}
\def\JMosPS{{\tscors J. Moscow Phys. Soc. }}
\def\IJMatP{{\tscors Int. J. Math. Phys  }}

\def\MPLA{{\tscors Modern Physics Letters A  }}
\def\CQG{{\tscors Class. Quantum Grav. }}

\def\d{\partial}

\def\sc{\scriptstyle}
\def\scc{\scriptscriptstyle}
\def\note{\advance\notenumber by 1 \footnote{$^{\the\notenumber}$}}
\def\ref#1{\medskip\everypar={\hangindent 2\parindent}#1}
\def\beginref{\begingroup
\bigskip
\leftline{\titolino References.}
\nobreak\noindent}
\def\endref{\par\endgroup}
\def\ra{\rightarrow}
\def\beginsection #1. #2.
{\bigskip
\leftline{\titolino #1. #2.}
\nobreak\noindent}

\nopagenumbers
\null
\vskip 5truemm
\rightline {DFTT 6/94}
\rightline {SISSA 25/94/A}
\rightline{February 1994}
\vskip 15truemm
\centerline{\titolino A SCHR\"ODINGER EQUATION FOR MINI UNIVERSES}
\vskip 15truemm
\centerline{\tsnorm Marco Cavagli\`a$^{(a),(d)}$,
Vittorio de Alfaro$^{(b),(d)}$ and Alexandre T. Filippov$^{(c)}$}
\bigskip
\centerline{$^{(a)}$\tscorsp Sissa - International School for Advanced
Studies,}
\smallskip
\centerline{\tscorsp Via Beirut 2-4, I-34013 Trieste, Italy.}
\bigskip
\centerline{$^{(b)}$\tscorsp Dipartimento di Fisica
Teorica dell'Universit\`a di Torino,}
\smallskip
\centerline{\tscorsp Via Giuria 1, I-10125 Torino, Italy.}
\bigskip
\centerline{$^{(c)}$\tscorsp Joint Institute for Nuclear Research}
\smallskip
\centerline{\tscorsp R-141980 Dubna, Moscow Region, RUSSIA.}
\bigskip
\centerline{$^{(d)}$\tscorsp INFN, Sezione di Torino, Italy.}
\vskip 15truemm
\centerline{\tsnorm ABSTRACT}
\begingroup\tsnorm\noindent
We discuss how to fix the gauge in the canonical treatment of
Lagrangians, with finite number of degrees of freedom, endowed with time
reparametrization invariance. The motion can then be described by an
effective Hamiltonian acting on the gauge shell canonical space. The
system is then suited for quantization. We apply this treatment to the
case of a Robertson--Walker metric interacting with zero modes of bosonic
fields and write a \S equation for the on--shell wave function.
\vfill
\leftline{\tsnorm PACS: 04.20.Fy, 04.60.Ds, 04.60.Kz, 98.80.Hw.\hfill}
\smallskip
\hrule
\noindent
\leftline{E-Mail: CAVAGLIA@TSMI19.SISSA.IT\hfill}
\leftline{E-Mail: VDA@TO.INFN.IT\hfill}
\leftline{E-Mail: FILIPPOV@THSUN1.JINR.DUBNA.SU\hfill}
\endgroup
\vfill
\eject
\footline{\hfill\folio\hfill}
\pageno=1
\baselineskip=2\normalbaselineskip
\beginsection 1. Introduction.
Models of mini universes are characterized by a finite (small) number of
gravitational and matter degrees of freedom. Now, a quantum theory for a
mini universe may be relevant to the initial stages of the universe if an
underlying field structure (strings, or any sort of quantum field
treatment) is not yet predominant; then a quantum treatment of a few
degrees of freedom could make sense [1,2]. The formulation of a
consistent quantum theory for these models constitutes a very interesting
problem.

The general covariance of General Relativity was at the basis of its very
original formulation; and later, different approaches have investigated
its canonical formulation. The introduction of the lapse function and
shift vector, that are Lagrange multipliers, goes back to 1962 [3]. The
investigations of the canonical structure of General Relativity have made
large progress   in the recent years together with the discussions about
the definition and meaning of time (see [4-6] and the contributions to
[7-9]). A general treatment of quantization of relativistic gauge systems
is discussed in [10].

In mini universe models there is a residual invariance under
reparametrization of time, similarly to what we find for particle
problems in special relativity, and the Wheeler-DeWitt (WDW) equation
[11,12] is the quantum expression of the correspondent constraint. Their
canonical treatment requires analysis of the gauge invariance and
reduction of the redundant degrees of freedom either by the BRST
formalism or, alternatively, by reducing the phase space by gauge fixing.
The book by Henneaux and Teitelboim is a guideline to the canonical
treatment of classical and quantum gauge systems [13].

In this paper our limited aim is to discuss their canonical structure and
the quantum formulation of mini universe models. In our examples we will
consider a Robertson-Walker (RW) universe, thus a single gravitational
degree of freedom, its scale factor $a$.

As for matter, aiming to have consistently radiation with the same
symmetry as the RW metric, we will introduce an SU(2) YM field and
consider its zero mode [14-16].  Of course, the presence of a YM field is
very welcome, as gauge fields constitute a fundamental ingredient of
matter. This mini universe has the right properties of a radiation
dominated universe: a separable WDW constraint, corresponding to the
classical property of radiation density scaling as $a^{-4}$.  A further
component of matter that does not spoil this simplicity is a conformal
scalar field, and we will introduce its zero mode for completeness.

The model under consideration does not contain, as it stands, interaction
terms producing inflation; inflation must be introduced by hand in the
Lagrangian in the form of a cosmological term and we will discuss briefly
a form of gauge fixing for this case.  We emphasize that the method we
are presenting is not restricted to these cases; indeed, the next step
will be to consider a minimally coupled scalar field. This is of great
interest as it introduces a coupling to gravitation that induces
inflation in a dynamical way. This model is more complex and will be
discussed elsewhere.

The dynamics of the mini universe is expressed by a Lagrangian
$$L=p_\mu \dot q_\mu-lC(q_{\mu}p_{\mu})~~~~~~\mu=0,...,N-1\eqno(1.1)$$
which displays gauge invariance under reparametrization of time (here we
reserve the word ``gauge'' to the case of time reparametrizations) and
has close similarity to the dynamics of the free relativistic particle or
of the relativistic harmonic oscillator.  $l$ is the Lagrange multiplier.
In order to eliminate redundancy and reduce the canonical degrees of
freedom the central point is to introduce a dynamical way of fixing the
gauge in the form of a relationship between canonical variables and time:
$$F(q,p,t)=0.\eqno(1.2)$$
The relationship (1.2), together with the constraint $C=0$, must allow to
eliminate one canonical degree of freedom as function of the remaining
ones and of time; thus time is connected to the original set of canonical
variables and the problem is reduced to gauge shell with 2$(N-1)$
canonical coordinates  (analogously to the unitary gauge in field
theories). Naturally the Poisson bracket $\{F,C\}$ must not vanish even
weakly (see for instance [13,17,18]), thus the set $(F,C)$ is second
class. The condition (1.2) must determine the Lagrange multiplier $l(t)$
(essentially the lapse function: $l=N/a$) as a function of time and in
general of canonical variables.  In particular in some cases we will
obtain the so called cosmic time gauge $N=1$, in others $N=a$, the
conformal time.  The gauge fixing function $F$ must be explicitly
dependent upon time since the constraint is not linear in coordinates and
momenta.

The reduction is implemented by a canonical transformation.  For
instance, if $F$ is of the form $F=p_0-f(q_{\mu},t)$ one can choose
$P_0=F,~Q_0=q_0$.  Then the gauge shell is obtained by imposing $P_0=0$
and fixing $Q_0$ from the constraint $C=0$.  One canonical degree of
freedom is thus eliminated.  The motion is reduced to canonical form in
$N-1$ degrees of freedom with an effective Hamiltonian (in general time
dependent) generating canonical motion on the gauge shell.  Eq.  (1.2) is
the fundamental gauge fixing equation that states how time is connected
to canonical variables. This implements a program of general relativity,
namely time as defined in terms of physical variables.

The basic criterion for the choice of the gauge fixing condition must be
the simplicity of the motion in the subspace of physical variables. Of
course the difficulty resides in finding a gauge that simplifies the
expression of the Hamiltonian on gauge shell (and must obey the
requirements about gauge fixing that we know from field theories). The
expression for $l(t)$ is a consequence of that choice and thus not of
primary interest in the procedure.

It is then natural to quantize the system in the reduced canonical space
(gauge shell) by writing a time dependent \S equation, or, what has the
same content, the Heisenberg equations.  In a given gauge, time is
connected to the canonical degree of freedom that has been eliminated by
the constraint and gauge fixing. This is the way to decide which degree
of freedom is better suited to play the role of time, whether the
gravitation degree of freedom or a matter variable connected to the YM
field (or to the conformal scalar).  Of course, reduction of the
gravitational degree of freedom to the role of time in a \S equation is
preferable, since its contribution in the constraint equation, for the
simple cases under consideration, has opposite sign to that of the
radiation degrees of freedom.  However, if one introduces other sorts of
matter, as for instance a scalar field, mixed terms will appear in the
constraint and the gauging off of gravitation is more complicate.  We
shall examine this case in a next investigation.

The wave function introduced in this way has the usual properties of
quantum mechanics, in particular a conserved current with positive
definite probability density exists, while the WDW equation is a gauge
constraint off-shell and  the WDW wave function is non physical.

The method used here has no difficulty (except complication) to be
implemented in the case of a set of first class constraints. Indeed one
then introduces corresponding gauge fixing terms in order to form  a set
of second class constraints and then proceeds to the elimination of the
redundant variables by the canonical transformation and establishes the
effective Hamiltonian on the gauge shell.
\beginsection 2. Canonical gauge fixing.
As we have said, our aim is to discuss the canonical formulation of mini
universe models in view of establishing the corresponding quantum theory.
 We use the ADM approach and in this paper we will treat the simple case
of a RW metric. Systems of this kind are endowed with time
reparametrization invariance. Other examples of systems with these
characteristics are the relativistic particle in Minkowsky space or the
relativistic harmonic oscillator.

In our case the first-class Hamiltonian must be zero (as is the case when
the canonical variables transform as scalars under reparametrization).
In this case the extended Hamiltonian will contain only the constraint
[13,17]:
$$H_{\rm E}=l(t)C(q_{\mu},p_{\mu}) \eqno(2.1)$$
where $l(t)$ is the Lagrange multiplier that imposes the constraint
$C=0$.

As a warming up exercise let us consider the relativistic massive free
particle. Its action is
$$S=\int dt\bigl(~-p_0 \dot x_0 + p_i \dot x_i~-~l(t)~C\bigr)\eqno(2.2)$$
where
$$C~\equiv~{{1}\over{2}}~({\bar p}^2+m^2-p_0^2)=0\eqno(2.3)$$
is the constraint term. In order to show how the gauge fixing works, let
us perform the following canonical transformation:
$$\eqalignno{&X_0~=~x_0-t~p_0,&(2.4a)\cr
&P_0~=~p_0.&(2.4b)\cr}$$

Let now take as gauge fixing condition
$$X_0~=~0.\eqno(2.5)$$

Of course the Poisson bracket $\{X_0,C\}$ is not vanishing weakly (it is
not a linear combination of $X_0$ and $C$). The use of the condition
(2.5) in the canonical equations of motion generated by (2.1) fixes the
Lagrange multiplier as follows:
$$l(t)~=~1.\eqno(2.6)$$
Using the gauge fixing (2.5) and the constraint equation (2.3) the
constrained action is
$$S_c~=~\int~ dt~ \biggl(p_i\dot x_i-{1\over 2}P_{0}^2\biggr)\eqno(2.7)$$
with $P_0^2$ given by (2.3) and (2.4). Thus, with this choice of gauge,
in the sector $p_i,x_i$ the motion is generated by the Hamiltonian
$$H_{\rm eff}~=~{1\over 2}P_0^2~=~{1\over 2}({\bar p}^2+m^2).\eqno(2.8)$$
Note that this Hamiltonian generates motion in the physical 6-dimensional
space $(x_i,p_i)$; this is equivalent to the motion in the sector
$p_i,x_i$ generated by the action (2.2) with the constraints (2.3) and
(2.5). Let us note that the canonical transformation (2.4) must involve
the time explicitly and that (2.5) can be thought of as the definition of
time in terms of canonical variables. We remark that in this canonical
gauge fixing the final goal is to obtain a simple expression for the
effective Hamiltonian; the expression (2.6) of the Lagrange multiplier
$l(t)$ is a consequence of the gauge fixing (2.5) and of the equations of
motion, and does not concern us too much.  We will keep this in mind when
dealing with mini universes in general relativity.

It is interesting to compare this result with a different gauge fixing
condition. Let us choose as time the 0-th component of $x_{\mu}$. To do
that, eq. $(2.4a)$ has to be substituted by
$$X_0~=~x_0-t.\eqno(2.4a')$$
Then by the same procedure one arrives to the square root Hamiltonian:
$$H_{\rm eff}~=~\sqrt{{\bar p}^2+m^2}.\eqno(2.9)$$
This simple example clarifies the method whose ultimate effectiveness
resides in the possibility of finding a simple canonical transformation
leading to a simple $H_{\rm eff}$.

After these preliminaries, let us go back to the general case (2.1) and
define a canonical transformation of the form
$$\eqalignno{&P_\mu=p_\mu-f_{\mu}(q_\nu;t),&(2.10a)\cr
&Q_\mu=q_\mu.&(2.10b)\cr}$$
(the role of the $p$'s and $q$'s may be interchanged of course). From the
Poisson identity $\{P_{\mu},P_{\nu}\}=0$ follows, at least locally, that
$f_{\mu}=\d/\d q_{\mu} f(q_{\nu},t)$. In the following we will denote
partial differentiation by the symbol ``$|$''. Then we have
$$p_\mu\dot q_\mu=P_\mu\dot Q_\mu-f_{\sc |t}(Q_\mu;t)+{{d}\over {dt}}
f(Q_\mu;t).\eqno(2.11)$$
Now let us introduce the gauge fixing condition in the form
$$P_0=0.\eqno(2.12)$$
Note that the Poisson bracket of the gauge fixing (2.12) and of the
constraint, $\{P_0,C\}$, must not be weakly zero [13]. The function $f$
must depend explicitly on $t$ in order that the procedure works, namely
to fix completely the Lagrange multiplier. Now the system has two
constraints of second class, $P_0=0$ and $C=0$. The variable $Q_0$
conjugate to $P_0$ is fixed by the constraint: $Q_0=Q_c(Q_i,P_i;t)$,
where $i=1,...,N-1$ and $Q_c$ is obtained from
$$\bigl[C\bigl(Q_0,Q_i,p_0=f_{\sc | Q_{\sc 0}}(Q_0,Q_i;t),
P_i+f_{\sc |Q_{\scc
i}}\bigr)\bigr]_{Q_0=Q_c}=0.\eqno(2.13)$$
The system is now on the ``gauge shell'', the $2(N-1)$-hypersurface in
the $2N$-phase space $(Q_\mu,P_\mu)$ defined by
$$P_0=0,~~~~~~~~~Q_0=Q_c(Q_i,P_i;t).\eqno(2.14)$$
The effective Lagrangian in the $(Q_i,P_i)$ sector is obtained from
(2.1) using  (2.11) and (2.14). Neglecting  a total derivative we have
$$L_{\rm eff}=L[P_0=0,Q_0=Q_c]=P_i\dot Q_i -
H_{\rm eff}(Q_i,P_i;t)\eqno(2.15)$$
where the effective Hamiltonian on the gauge shell is
$$H_{\rm eff}(Q_i,P_i;t)=
f_{|t}\bigl(Q_c,Q_i;t\bigr).\eqno(2.16)$$
The effective Hamiltonian is in general time dependent. One can check
that the canonical equations
$$\eqalignno{&\dot Q_i~=~{\d H_{\rm eff}\over\d
P_i},&(2.17a)\cr
&\dot P_i~=~-{\d H_{\rm eff}\over\d Q_i}&(2.17b)\cr}$$
are equivalent to the original canonical system using (2.13) and (2.19)
(see below).

Finally, for consistency with the gauge fixing condition (2.12) we
require also
$$\dot P_0=0\eqno(2.18)$$
that determines the expression for the Lagrange multiplier:
$$l(t)=-\biggl[{f_{\sc |t q_{\scc 0}}\over C_{\sc |q_{\scc 0}}
+f_{\sc |q_{\scc 0}q_{\scc \mu}}C_{\sc
|p_{\scc\mu}}}\biggr]_{{\sc q_0=Q_c}\atop{\sc p_{\scc\mu}=
f_{\scc \mu}}}. \eqno(2.19)$$
In general the expression (2.19) can be complicated, however this does
not necessarily concern us: the important fact is that the gauge fixing
of the form (2.12) determines $l(t)$.

The problem of the motion is thus reduced to the task of determining the
gauge function $f(q_\mu;t)$ so that $H_{\rm eff}$ is as simple as
possible. In the next section we will discuss the gauge fixing for RW
universe in interaction with radiation.

In conclusion, let us remark that time is determined by the gauge fixing
condition to be a function of the variables $q_{\mu}$ and $p_{\mu}$ from
the condition $P_0=0$.  As we shall see, often $t$ turns out to be a
function only of a couple of canonical variables $p_0,q_0$, thus a single
degree of freedom defines the time in that gauge.	This is an
interesting situation.  A particular situation occurs when $t$ depends
only on $q_0$ or $p_0$.  This allows identification of time in that gauge
with a canonical coordinate of immediate physical relevance, and of
course we have in mind in the gravitational case the identification of
$t$ with the scale factor $a$ of the RW universe.  We shall see that,
while this is in general possible, this is not the most advisable choice
from the point of view of simplicity of $H_{\rm eff}$.  Indeed, the main
criterion for the choice of the function $f$ is, as we have said, to
obtain a simple $H_{\rm eff}$.	With the identification of time with a
degree of freedom $H_{\rm eff}$ is in general time dependent.

In principle there is no great difficulty to generalize this method.
After all, the four constraints in a general invariant theory are of
first class and the procedure exposed here can be repeated.  We stress,
the problem is the determination of suitable canonical transformations
(i.e. gauge fixing identities) so that the corresponding effective
Hamiltonian is simple and possibly time independent.
\beginsection 3. Simple examples of gauge fixing.
We will apply the ideas exposed above to some simple cases of relevance
to the discussion of mini universe models with RW or de Sitter spacetime
coupled to zero modes of different fields. Let the constraint have the
form
$$C\equiv {1\over 2}p_0^2+V(q_0)-H_1(q_\mu,p_i)=0\eqno(3.1)$$
We may identify $q_0$ as the gravitational degree of freedom, however
this is not always needed.

Let us first assume that the constraint is separable and the 0-th degree
of freedom has the form of a harmonic oscillator: this happens, for
instance, when we deal with a conformal scalar field or with a RW closed
metric. So we have
$$\eqalignno{&V(q_0)={1\over 2}q_0^2,&(3.2a)\cr
&H_1\equiv H_1(q_i,p_i).&(3.2b)\cr}$$
We choose the canonical transformation (2.10) as
$$\eqalignno{&P_0=p_0+q_0\hbox{ctg}~t,&(3.3a)\cr
&Q_0=q_0&(3.3b)\cr}$$
and the gauge fixing condition is of the form
$$p_0+q_0~\hbox{ctg}~t=0.\eqno(3.4)$$
Gauge fixing conditions of this sort were introduced in [19] in the
context of a gauge approach to systems of many relativistic particles. By
the method exposed in section 2 one obtains a very interesting result for
the effective Hamiltonian on the gauge shell:
$$H_{\rm eff}=~H_1.\eqno(3.5)$$
The simplicity of (3.5) shows the interest of the gauge fixing (3.4).
$l(t)$ is fixed:
$$l(t)=-1.\eqno(3.6)$$
Thus in this problem the natural choice of the time is
$\hbox{arctg}~q_0/p_0$.

One would intuitively like to identify $q_0$ as time. Let us see what
happens if one assumes gauge fixing conditions of the form
\bigskip
\line{\hfill$p_0=\sqrt{2}~t,$\hbox to 30truemm{\hfill or\hfill}
$q_0=-\sqrt{2}~t.$\hfill (3.7)}
\bigskip\noindent
Then respectively
$$l(t)=-{{\sqrt{2}}\over {q_0}},~~~~l(t)=-{{\sqrt{2}}\over {p_0}}.
\eqno(3.8)$$
The effective Hamiltonian is time dependent:
$$H_{\rm eff}=~2\sqrt{H_1-t^2}.\eqno(3.9)$$
The positive definiteness of the operator under square root implies that
the support of $q_0$ or $p_0$ is restricted, in agreement with the
general properties of the oscillatory motion in $q_0$, $p_0$. One sees
the advantage of the gauge choice (3.4) since in that case $H_{\rm eff}$
is  independent of time.

Let us end this part dedicated to the oscillator by noticing that of
course a non compact oscillator requires hyperbolic functions in place of
circular.

In mini universe models we shall encounter a quartic potential, for
instance for the zero mode of the Yang--Mills field and of course in the
case of a cosmological term. So let us discuss the gauge fixing in case
$V(q_0)$ of (3.1) has the form
$$V(q_0)=kq_0^2-\lambda q_0^4\eqno(3.10)$$
and $(3.2b)$ holds. In this case it is convenient to choose the canonical
transformation as
$$\eqalignno{&P_0=p_0-\sqrt{2\lambda}\bigl(q_0^2+g(t)\bigr),&(3.11a)\cr
&Q_0=q_0&(3.11b)\cr}$$
where $g(t)=t^2-k/2\lambda$. The gauge fixing condition (2.12) is
$$p_0=\sqrt{2\lambda}\bigl(q_0^2+g(t)\bigr)\eqno(3.12)$$
which together with (2.18) allows to fix the Lagrange multiplier
$l(t)$:
$$l(t)=-{1\over\sqrt{2\lambda}tq_0}.\eqno(3.13)$$
We may now compute the effective Hamiltonian in the physical degrees of
freedom:
$$H_{\rm eff}~=\sqrt{2\lambda}~\dot g(t)Q_c.\eqno(3.14)$$
In eq. (3.14) $Q_c$ must be obtained from the constraint as in (2.13),
$$Q_c^2={1\over k+2\lambda g(t)}(H_1-\lambda g^2(t))\eqno(3.15)$$
and thus
$$H_{\rm eff}=2\sqrt{H_1-\lambda(t^2-k/2\lambda)^2}\eqno(3.16)$$
It is evident that the final form of $H_{\rm eff}$ is not simple and is
time dependent.  What is important is that time is essentially fixed by
(3.12) as a function of $p_0$ and $q_0$.

Now let us give some hints about the case of a general potential in (3.1)
(note that often this procedure is not the best to follow, as it
happens, for instance, in the case just discussed).

Eq. (2.13) that defines  $Q_c(q_i,p_i;t)$ is now
$$\biggl[{1\over 2}\bigl(f_{\sc|Q_{\scc 0}}(Q_0,Q_i;t)\bigr)^2+
V(Q_0)-H_1(Q_0,Q_i,P_i+
f_{\sc |Q_{\scc i}})\biggr]_{Q_0=Q_c}=0.\eqno(3.17)$$
Remembering the form (2.16) of $H_{\rm eff}$ let us choose $f\equiv
f(Q_0;t)$; one needs to connect  $f_{\sc|Q_{\scc 0}}$ to $f_{|t}$ and we
may proceed for instance as follows. Set
$${1\over 2 }\bigl(f_{|q}(q;t)\bigr)^2+V(q)~=~g(t)~F(q)\eqno(3.18)$$
where $g(t)$ and $F(q)$ arbitrary functions of $t$ and of $q$
respectively. Then
$$f(Q_0;t)~=~\sqrt{2}\int^{Q_0} dq~\bigl[H_1(q,Q_i,P_i)-
V(q)\bigr]^{1/2}\eqno (3.19)$$
and the effective Hamiltonian has the expression
$$H_{\rm eff} = {{\dot g} \over{\sqrt{2}}} \int^{Q_c} dq\sqrt{F(q)}
{}~\biggl(g(t)-{{V(q)}\over{F(q)}} \biggr)^{-1/2}.\eqno(3.20)$$
$F(q)$ may be chosen so as to simplify the expression (3.19): for
instance, if $V(q_0)$ is of the form $q_0^2v(q_0)$ with $v(q_0)$ a
polynomial, then $F=q^2$ is a suitable choice. Of course if the potential
is particularly simple one may set
$$F(q)~=~V(q).\eqno(3.21)$$
Then
$$H_{\rm eff}~=~\sqrt{2}{{d}\over{dt}}(g-1)^{1/2} \int^{Q_c} dq
\sqrt{V(q)}.\eqno(3.22)$$
Note that  $H_{\rm eff}$ is in general time dependent also because of
$Q_c$. Finally, the Lagrange multiplier is
$$l(t)=\biggl[-{f_{\sc |tq_{\scc 0}}\over V'(q_0) +f_{\sc
|q_{\scc 0}}f_{\sc |q_{\scc 0}q_{\scc 0}}-H_{\sc 1|q_{\scc 0}}
(q_0,q_i,f_{\sc |q_{\scc i}})}\biggr]_{q_0=Q_c}.\eqno (3.23)$$
The previous results about the harmonic oscillator can be obtained from
these general formulae choosing
$$f(q_0;t)=-{1\over 2}q_0^2\hbox{ctg}~t\eqno(3.24)$$

In the end let us apply the method discussed above to a constraint of the
form
$$C={1\over 2}p_0^2-{1\over 2q_0^2}p_1^2-q_0^4
V_1(q_1)+V_0(q_0).\eqno(3.25)$$
This case corresponds to the zero mode of a scalar field minimally
coupled to gravity. Now $Q_c(q_i,p_i;t)$ is defined by
$$\biggl[{1\over 2}\bigl(f_{\sc |Q_{\scc 0}})^2-Q_0^4 V_1(Q_1)+
V_0(Q_0)-(P_1+f_{\sc |Q_{\scc 1}})^2{1\over
2Q_0^2}\biggr]_{Q_0=Q_c}=0.\eqno(3.26)$$
Choosing $f\equiv f(Q_0;t)$ and setting
$${1\over 2}\bigl(f_{|q}(q;t)\bigr)^2+
V_0(q)\equiv q^4 g(t)\eqno(3.27)$$
we obtain the effective Hamiltonian
$$H_{\rm eff}={\dot g \over\sqrt{2}} \int^{Q_c} dq q^2
{}~\biggl(g(t)-{V_0(q)\over q^4} \biggr)^{-1/2}\eqno(3.28)$$
where now
$$Q_c=\biggl({P_1^2\over 2\bigl(g(t)-
V_1(Q_1)\bigr)}\biggr)^{1/6}.\eqno(3.29)$$
In particular, if
$$V_0(q_0)=q_0^4\eqno(3.30)$$
(3.28) becomes
$$H_{\rm eff}=\pm{1\over 3}{d\over
dt}\sqrt{g-1}{P_1\over\bigl(g(t)-V_1(Q_1)\bigr)^{1/2}}.\eqno(3.31)$$
Analogously to the cases discussed before, a suitable choice for $g(t)$
can simplify the effective Hamiltonian. Clearly, (3.31) depends
explicitly on time.

Finally the Lagrange multiplier is
$$l(t)=\biggl[-{q_0f_{\sc |tq_{\scc0}}\over 2 V_0+
q_0 V_0'-6q_0^4 V_1+f_{\sc
|q_{\scc 0}}\bigl(f_{\sc |q_{\scc 0}}+q_0f_{\sc |q_{\scc 0}q_{\scc 0}}
\bigr)}\biggr]_{q_0=Q_c}.\eqno (3.32)$$
We will not give here a complete discussion of the gauge reduction of
(3.25) because the coupling term $q_0^4 V_1(q_1)$ makes it difficult to
obtain a simple effective Hamiltonian. Let us discuss the simple case
$V_1(q_1)=0$ and $V_0(q_0)=q_0^2$. In this case it is convenient to
choose
$$f(q_0;t)={1\over\sqrt{2}}q_0^2\sinh{t}\eqno(3.33)$$
Using (3.26) we have that $Q_c(Q_i,P_i;t)$ is defined by
$$Q_c^2=\pm{1\over\sqrt{2}}{P_1\over\cosh{t}}\eqno(3.34)$$
and the effective Hamiltonian (2.16) can be written
$$H_{\rm eff}=\pm{1\over 2} P_1\eqno(3.35)$$
A surprising feature of (3.31) and (3.35) is that the effective
Hamiltonian is linear in $P_1$. In the next section we will apply all
these considerations to minisuperspace models.
\beginsection 4. Minisuperspace models.
Let us consider the action for gravity minimally coupled to SU(2)
Yang-Mills (YM) field $A$ and a conformal scalar (CS) field $\varphi$.
The reason for this choice is that for these fields the WDW equation
decouples, a fact directly related to the property that the classical
energy density scales as $a^{-4}$. The action is the sum of three terms:
$$I=I_{\rm GR}+I_{\rm YM}+I_{\rm CS}.\eqno(4.1)$$
Here $I_{\rm GR}$ represents the Einstein-Hilbert action of the
gravitational field
$$I_{\rm GR}=-\int_\Omega
d^4x\sqrt{-g}\bigl(R+2\Lambda\bigr)+2\int_\Omega
d^3x\sqrt{h}~{\bf K}\eqno(4.2a)$$
where $R$ is the scalar curvature tensor, $\Lambda$ the cosmological
constant and ${\bf K}$ the extrinsic curvature of the manifold $\Omega$.
We use definitions as in Landau - Lifshitz and we have put $16\pi
G\equiv~L_p^2~\equiv~M_p^{-2}=1$ thus measuring all dimensional
quantities in these units. The second and third term in (4.1), $I_{\rm
YM}$ and $I_{\rm CS}$, represent respectively the YM and the CS field
actions
$$\eqalignno{I_{\rm YM}&={1\over 2}\int F\wedge{}^*F,&(4.2b)\cr
I_{\rm SC}&={1\over 2}\int
d^4x\sqrt{g}\biggl(\partial_\mu\varphi\partial^\mu\varphi+{1\over
6}R\varphi^2\biggr)&(4.2c)\cr}$$
where $F=dA+A\wedge A$ is the field strength 2-form and we have set = 1
the gauge coupling constant.

We shall study spacetimes of topology $R\times H$, where $H$ is a
homogeneus and isotropic three-surface. Hence we shall write
$$ds^2=N^2(t)dt^2-a^2(t)\omega^p\otimes\omega^p
\eqno(4.3)$$
where $\omega^p$ are the 1-forms invariant under translations in space
and $N(t)$ is the lapse function. The cosmic time     corresponds to
$N=1$ and the conformal time   to $N=a$. In the cases we are going to
discuss, the $\omega^p$'s will satisfy the Maurer-Cartan structure
equations:
$$d\omega^p={k\over 2}\epsilon_{pqr}\omega^q\wedge\omega^r\eqno(4.4)$$
where $k=0,1$ and thus the line element (4.3) describes a flat or a
closed Friedmann-Robertson-Walker universe. When $k=1$, (4.3) has the
SU(2)$_L$ $\times$ SU(2)$_R$ group of isometries.
Using (4.3) and integrating over the spatial variables, the space density
of the gravitational action $(4.2a)$ can be written
$$S_{\rm GR}=6\int dt\biggl(-{a\dot a^2\over N}
+kNa-{\Lambda\over 3} Na^3\biggr).\eqno(4.5)$$
Introducing the conjugate momentum
$$p_a~=~-12{a\dot a\over N}\eqno(4.6)$$
(4.5) can be cast in the form
$$S_{\rm GR}=\int dt \biggl(p_a\dot a~+~{N\over a}~
H_{\rm GR}\biggr).\eqno(4.7)$$
Here $H_{\rm GR}$ is
$$H_{\rm GR}~=~{1\over 12}\biggl({1\over 2}p^2_a+V_a(a)\biggr)
\eqno(4.8)$$
where
$$V_a(a)=72a^2\biggl(k-{\Lambda\over 3}a^2\biggr).\eqno(4.9)$$
%
%
Let us now introduce a YM field configuration with the same symmetry as
the metric.  We shall use a YM group SU(2) for simplicity and (in line
with the metric) a form of the YM field with a single degree of freedom
that was proposed in [14] (see also [16]); the case of a more general
group has been investigated in [15]. The form of the field, written in
the dreibein cotangent space, is
$$A={i\over 2}\xi(t)\sigma_p\omega^p.\eqno(4.10)$$
With the definition (4.10) $A$ is evidently left- invariant; it is also
right- invariant up to a gauge transformation [14,20]. $\xi(t)$ is the
single degree of freedom.

The field strength $F$ is:
$$F={i\over 2}\sigma_p\dot \xi dt \wedge \omega^p~ +
{}~{i\over 4}\sigma_r\epsilon_{rpq}\xi(k-\xi) \omega^q \wedge \omega^r.
\eqno (4.11)$$
Thus the action of the YM field becomes
$$S_{\rm YM}={3\over 2}\int dt
\biggl({a\over N}\dot \xi^2-{N\over a}\xi^2(k-\xi)^2\biggr)
\eqno(4.12)$$
and introducing the conjugate momentum
$$p_\xi=3{a\over N}\dot \xi,\eqno(4.13)$$
(4.12) can be cast in the form
$$S_{\rm YM}=\int dt\biggl( p_\xi\dot \xi-
{{N}\over{a}}~H_{\rm YM}\biggr)\eqno(4.14)$$
where $H_{\rm YM}$ is
$$H_{\rm YM}={1\over 3}\biggl({1\over
2}p^2_{\xi}+V_\xi(\xi)\biggr)\eqno(4.15)$$
and
$$V_\xi(\xi)={9\over 2}\xi^2(k-\xi)^2.\eqno(4.16)$$
Now, let us discuss the CS field. Using (4.3) the CS field action becomes
$$S_{\rm CS}={1\over 2}\int d t\biggl({a\over N}(a\dot \varphi+\dot
a\varphi)^2-kNa\varphi^2\biggr).\eqno(4.17)$$
Defining the rescaled scalar field $\chi=\varphi a$, (4.17) can be cast
in the form
$$S_{\rm CS}={1\over 2}\int dt\biggl({a\over N}\dot\chi^2-k{N\over
a}\chi^2\biggr).\eqno(4.18)$$
Introducing as in the previous cases the conjugate momentum
$$p_\chi={a\over N}\dot\chi,\eqno(4.19)$$
(4.18) becomes
$$S_{\rm CS}=\int dt \biggl(p_\chi\dot\chi-{{N}\over{a}}
{}~H_{\rm CS}\biggr)\eqno(4.20)$$
where
$$H_{\rm CS}={1\over 2}\bigl(p_\chi^2+k\chi^2\bigr).\eqno(4.21)$$
In the conformal gauge, when $k=1$, the Hamiltonian (4.21) describes a
one-dimensional harmonic oscillator in the $\chi$ variable.

The classical Friedmann - Einstein equation of motion is the constraint
equation
$$H_{\rm YM}+H_{\rm CS}=H_{\rm GR}.\eqno(4.22)$$
{}From the motion equations for the YM and the CS fields it is easy to
obtain
$$\eqalignno{&\biggl({a\over N}\dot\xi\biggr)^2+\xi^2(k-\xi)^2=K^2_{\rm
YM}&(4.23a)\cr
&\biggl({a\over N}\dot\chi\biggr)^2+k\chi^2=K^2_{\rm CS}&(4.23b)\cr}$$
where $K_{\rm YM}$ and $K_{\rm CS}$ are independent of $t$; using (4.23),
(4.22) becomes
$${a^2\dot a^2\over N^2}+\biggl(ka^2-{\Lambda\over 3}a^4\biggr)={1\over
12}K^2\eqno(4.24)$$
where
$$K^2=3K_{\rm YM}^2+K_{\rm CS}^2.\eqno(4.25)$$
In the cosmic gauge, $N=1$, this reads
$${\dot a^2\over a^2}+{k\over a^2}={\Lambda\over 3}+{1\over 12}
{K^2\over a^4}.\eqno(4.26)$$
We see that, as due to radiation, the energy density scales as $a^{-4}$.
The meaning of $K^2$ is obvious:
$$K^2=2\rho(a=1).\eqno(4.27)$$
$K^2$ is thus the energy density of the radiation in a RW universe with
scale factor $a$ of one Planck length.

Let us turn to gauge fixing and writing a \S equation. The complete
action of our radiation filled mini universe is
$$S=\int dt\bigl( p_a\dot a + p_{\xi}\dot\xi + p_{\chi}\dot\chi
{}~ - l(t)~ H_{\rm WDW}\bigr)\eqno(4.28)$$
where
$$H_{\rm WDW}~=~H_{\rm YM}~+~H_{\rm CS}~-~H_{\rm GR}\eqno(4.29)$$
and
$$l(t)~=~{{N(t)}\over{a}}.\eqno(4.30)$$

Now we see that in this academic model we may choose the gauge fixing in
essentially different ways. A very spontaneous and physically appealing
choice consists in using the gravitational degree of freedom as connected
to time. Let us first discuss the case $\Lambda=0$, i.e. absence of the
cosmological term, and a closed universe, $k=1$. Then  $H_{\rm GR}$ is a
harmonic oscillator and for a closed universe we use the gauge fixing
identity of the form (3.4),
$$p_a~=-12~a~\hbox{ctg}~t.\eqno(4.31)$$
As a consequence
$$l(t)~=~1.\eqno(4.32)$$
This is a conformal time gauge, $N=a$.

With this choice of time, the Hamiltonian is time independent; the \S
equation on the gauge shell takes the form
$$i~{{\d }\over{\d t}}~\Psi(\xi,\chi;t)~=~\bigl(~H_{\rm CS}~+~
H_{\rm YM}~\bigr) \Psi(\xi,\chi;t).\eqno(4.33)$$

Let us note that the classical Friedmann equation (4.24) for $k=1$, in
this gauge and with $\Lambda=0$, reads
$$\dot a^2 + a^2~=~{{K^2}\over{12}}~\equiv~a_{\rm M}^2\eqno(4.34)$$
and a solution for the classical motion is
$$a~=~a_{\rm M}~\sin{t}\eqno(4.35)$$
{}From (4.6) we have
$$p_a~=~- 12 \dot a\eqno(4.36)$$
in agreement with the gauge condition (4.31). With the present definition
of time, $t=\hbox{arctg}~a/\dot a$, the region $0 \leq t \leq \pi/2$ maps
the expanding phase of the closed universe. Boundary conditions are of
course expressed in the form
$$\Psi(\chi,\xi;t_0)~=~f(\chi,\xi).\eqno(4.37)$$
Eq. (4.33) expresses the evolution of the configuration in $t$ and
represents as usual in quantum mechanics the correlation amplitude for
the different components of matter ($\chi,~\xi$ in the present case) in
the universe.

Let us point out that a gauge choice identifying $a$ with time,
$$a={|t|\over\sqrt{6}}\eqno(4.38)$$
leads to a time dependent Hamiltonian
$$H_{\rm eff}=2\sqrt{H_{\rm YM}+H_{\rm CS}-t^2}.\eqno(4.39)$$

The real problem comes with a cosmological term. In that case $V_a$ has
the form (4.9) and we then use the form (3.20) with $g\ra 12g$ for
$H_{\rm eff}$ with the choice $F(a)=a^2$. Then
$$\eqalign{H_{\rm eff}&=\sqrt{6}\dot g
\int da~ a \bigl(g-6k+2\Lambda a^2\bigr)^{-1/2}\cr
&={1\over 2\sqrt{2}\lambda}{{\dot g}\over{g^{1/2}}}
\bigl( g(g-6k)+2\Lambda(H_{\rm YM}+H_{\rm CS})
\bigr)^{1/2}.\cr}\eqno(4.40)$$
Let us now discuss an interesting different gauge fixing for the present,
academic, case. We could very easily choose the time so as to be
connected to the CS degree of freedom, since its Hamiltonian is a simple
harmonic oscillator. This leads to ambiguities, already present in the
classical discussions of the WDW equation, about self -- adjointness of
the quantum Hamiltonian $H_{\rm GR}$, and about the differential
representation of the (now operator) $p_a$. It is interesting to examine
this case. Let us write the gauge fixing condition as
$$p_{\chi}~=~\chi~ctg~t.\eqno(4.41)$$
Again this is a conformal time gauge, $l(t)=1$. With this choice of time
the \S equation takes the form
$$i~{{\d }\over{\d t}}~\Psi(a,\xi;t)~=~ \bigl(~ H_{\rm YM}~-~H_{\rm
GR}~\bigr) \Psi(a,\xi;t).\eqno(4.42)$$
This identification of time connects the latter to a physically unclear
entity as the CS field. However, this form is suited to discuss
correlation between $a$ and $\xi$. We will see that we may draw some
consequences in agreement with the correspondence principle. Since $t$ is
physically not well defined, let us direct our attention to stationary
states. With
$$\Psi(a,\xi;t)~=~ e^{-iEt}~\psi(a,\xi)\eqno(4.43)$$
we obtain the stationary \S equation
$$\bigl( H_{\rm YM}~-~H_{\rm GR} \bigr)~\psi(a,\xi)~=~E~ \psi(a,\xi).
\eqno(4.44)$$

This equation is similar to the WDW equation for pure gravity + YM (see
[16] and also [21]), the difference being  the presence of the term
$E\psi$. The separability of the equation is the quantum form of the
request that the density scales like $a^{-4}$. Let us write
$$\Psi(a,\xi)=\psi(a)\eta(\xi).\eqno(4.45)$$
The equation for the YM wave function is
$$ {{1}\over{3}}~ \biggl({1\over
2}p^2_\xi+V_\xi(\xi)\biggr)\eta(\xi)=E_{\rm YM}~\eta(\xi).\eqno(4.46)$$
As to boundary conditions, we remark that if we assume that the wave
function tends to zero for large $|\xi|$, then eigenvalues are quantized.

Now let us turn to the gravitational degree of freedom again in the case
of a closed universe. We have
$${{1}\over{12}}~  \biggl({1\over 2}p^2_a+V_a(a)~\biggr)\psi_n(a)=
E^g_n\psi_n(a).\eqno(4.47)$$
The gravitational degree of freedom is a harmonic oscillator. It is then
natural to set the boundary condition at $a\rightarrow \infty$ by asking
the square integrability of the wave function (for this suggestion see
e.g. [16,22-24]). About the condition at $a=0$, if we ask that
$\psi\rightarrow 0$, then $p_a=-id/da$  is formally Hermitean and
$H_{\rm GR}$ is self-adjoint. Let us accept for the moment these boundary
conditions. Then the spectrum is given by the odd part of a harmonic
oscillator ($n_g$ odd)
$$E^g_n={1\over 2}+n_g\eqno(4.48)$$
and the wave functions are harmonic oscillator ones [16]. The eigenvalues
of the two degrees of freedom are connected by
$$E^g_n~=~E^{\rm YM}_n~ - ~E.\eqno(4.49)$$
Let us now consider the correspondence principle with the classical
gravitational motion for large oscillator quantum numbers $n_g$ in the
gauge $l(t)=1$ [16]. We interpret of course, as we are led by the \S
equation, $\vert \psi\vert^2$ as the probability density for the value
$a$ for the scale factor, to be compared through the correspondence
principle to the classical probability density in the conformal time
gauge distribution of the physical quantity $a$ for an ensemble of
trajectories. Now this is inversely proportional to the speed of $a$ in
that time gauge. Thus (probability not normalized)
$$P_{cl}(t)~=~{{1}\over{da/dt}}.\eqno(4.50)$$
{}From the classical equation of motion (4.34) we have in the conformal
time gauge
$$P_{cl}(t)~=~
{{1}\over{\sqrt{a_{\rm M}^2-a^2}}}.\eqno(4.51)$$
Now for the harmonic oscillator
$$\Sigma_{av}~\vert\psi_n(a)\vert^2~\ra~
{{1}\over{\sqrt{a_{\rm M}^2-a^2}}}.\eqno(4.52)$$
So from the boundary conditions chosen it follows that the correspondence
principle works properly in the conformal gauge $N=a$ as noted in [16].

The extension of the discussion of the correspondence principle to the
case of a universe  with a cosmological term can be carried on along the
lines shown in [16].

To conclude this section, let us explore briefly the case of gravity
minimally coupled to a scalar field $\phi$. Its action is
$$I_{\rm MCS}=\int d^4x\sqrt g\biggl[{1\over 2}\partial_\mu\phi
\partial^\mu\phi-V(\phi)\biggr]\eqno(4.53)$$
and the complete space density action in minisuperspace reduces to
$$S=\int dt\bigl(p_a\dot a + p_{\phi}\dot\phi
{}~ - l(t)~(H_{\rm MCS}-H_{\rm GR}\bigr)\eqno(4.54)$$
where
$$H_{\rm MCS}=\biggl({1\over 2}{p_\phi^2\over
a^2}+a^4V(\phi)\biggr).\eqno(4.55)$$
Let us put, for simplicity, $V(\phi)=0$ and consider a closed universe
$k=1$. Note that analogous results hold for the flat case $k=0$. Using
the linear gauge fixing
$$p_a-12a\sinh{(t/\sqrt{3})}=0,\eqno(4.56)$$
we obtain the effective Hamiltonian
$$H_{\rm eff}=\pm p_\phi.\eqno(4.57)$$
Note that we are essentially in the conformal gauge, in fact from (4.56)
we have
$$l={1 \over2\sqrt{3}\cosh{(t/\sqrt{3})}}.\eqno(4.58)$$
Then the \S equation is
$$\biggl({\d\over\d t}-(\pm){\d\over\d\phi}\biggr)\Psi=0\eqno(4.59)$$
which has the general solution
$$\Psi=f(\phi \pm t)\eqno(4.60)$$
\beginsection 5. Conclusions.
The classical equations of motion for mini universes are endowed with a
residual time reparametrization invariance. The ensuing constraint, upon
quantization, becomes the WDW equation
$$H_{WDW}~\Psi_{WDW}~=~0.\eqno(5.1)$$

Now this equation, fundamental as it is, contains well known ambiguities
about which much has been written: absence of time, absence of conserved
current, choice of boundary conditions, interpretation of the WDW wave
function and normalization, and so on.

We have considered the classical Lagrangian for mini universes at its
face value and have implemented the procedure of gauge fixing in the
canonical scheme. Note that this procedure amounts to a definition of
time in terms of canonical coordinates. The choice of the gauge is in
general a fine art and there is no a priori rule apart from the final
simplicity of the effective Hamiltonian on gauge shell. This is actually
the case also in gauge field theories.

Once the classical motion has been reduced to the gauge shell, the system
can be quantized. Then one obtains the \S equation for a mini universe
and the wave function has the usual properties of quantum mechanics:
there is time, there is a conserved current and a positive density, thus
the interpretation of the wave function is the usual one (we do not enter
into the problem of the meaning of quantum mechanics as applied to the
entire universe), and finally there is a $i~\d/\d t$ in the equation.

Our treatment follows thus the lines of rational mechanics. In
particular, we have not discussed here whether, in the case of a closed
universe, time is bounded and about consequences.

Recently, possible ways to quantize the gravitational field by solving
the contraints classically and then quantizing the reduced system have
been discussed [25-29]. We believe this is the right direction.
\vfill\eject
\beginref

\ref [1] C.W. Misner, \PR {\bf 186}, (1969) 1319.

\ref [2] C.W. Misner, in {\tscors Magic Without Magic: John Archibald
Wheeler, a Collection of Essays in Honour of His 60th Birthday}, ed. by
J. Klauder, Freeman, San Francisco, 1972.

\ref [3] R. Arnowitt, S. Deser and C.W. Misner, {\tscors The Dynamics
of General Relativity}, in {\tscors Gravitation: An Introduction to
Current Research}, Wiley, New York, 1962.

\ref [4] J. B. Barbour, \PRD {\bf 47}, (1993) 5422.

\ref [5] C. Rovelli, \PRD {\bf 43}, (1991) 442 and references therein.

\ref [6] C.J. Isham, Lecture given at the WE-Heraeus-Seminar {\tscors
The Canonical Formalism in Classical and Quantum General Relativity},
Bad Honnef, Germany, September 1993; Preprint Imperial/TP/93-94/1.

\ref [7] {\tscors Gravitation: A Banff Summer Institute}, Banff Center,
Canada, Aug. 12-25, 1990 ed. by R. Mann and P. Wesson; World Scientific,
Singapore, 1991.

\ref [8] A. Ashtekar, Lectures delivered at the {\tscors 1992 Les Houches
Summer School on Gravitation and Quantization}, Les Houches, France.

\ref [9] {\tscors Conceptual Problems of Quantum Gravity}, Proceedings
of the 1988 Osgood Hill Conference, ed. by A. Ashtekar and J. Stachel,
North Andover, Massachusetts, Boston, 1988.

\ref [10] P. H\'aj\'\i \v cek and K.V. Kucha\v r, \PRD {\bf 41}, (1990)
1091.

\ref [11] B. DeWitt, \PR {\bf 160}, (1967) 1113.

\ref [12] J.A. Wheeler, in: {\tscors Battelle Rencontres} eds. C. DeWitt
and J.A. Wheeler, Benjamin, New York, 1968.

\ref [13] M. Henneaux and C. Teitelboim, {\tscors Quantization of Gauge
Systems}, Princeton , New Jersey, 1992.

\ref [14] M. Henneaux, \JMP {\bf 23}, (1982) 830.

\ref [15] O. Bertolami, J.M. Mour\~ao, R.F. Picken and I.P. Volobujev,
\IJMatP {\bf A23}, (1991) 4149.

\ref [16] M. Cavagli\`a and V. de Alfaro, to appear in \MPLA.

\ref [17] A. Hanson, T. Regge and C. Teitelboim, {\tscors Constrained
Hamiltonian Systems}, Accademia Nazionale dei Lincei, Roma, 1976.

\ref [18] P.A.M. Dirac, {\tscors Lectures on Quantum Mechanics}, Belfer
Graduate School of Science, Yeshiva University, New York, 1964.

\ref [19] A.T. Filippov, \MPLA {\bf 4}, (1989) 463.

\ref [20] D.V. Galt'sov and M.S. Volkov, \PLB {\bf 256}, (1991) 17.

\ref [21] O. Bertolami and J.M. Mour\~ao, \CQG {\bf 8}, (1991) 1271.

\ref [22] H.D. Conradi and H.D. Zeh, \PLA {\bf 154} (1991) 321.

\ref [23] L.J. Garay, \PRD {\bf 48}, (1993) 1710.

\ref [24] J. H. Kung, {\tscors Quantization of Closed Minisuperspace
Models as Bound States}, PRINT-93-0196, (Harvard-Smithsonian Center for
Astrophysics), Feb. 1993, HEP-TH/9302016.

\ref [25] V. Pervushin and T. Towmasjan, \JMosPS {\bf 3}, (1993) 1.

\ref [26] R.P. Woodard, \CQG {\bf 10}, (1993) 483.

\ref [27] J.A. Rubio and R.P. Woodard, {\tscors Reduced Hamiltonians},
UFIFT-HEP-93-20, (Florida U.), Nov. 1993, GR-QC/9312020.

\ref [28] S. Carlip, \CQG {\bf 11}, (1994) 31; \PRD {\bf 42}, (1990)
2467.

\ref [29] G. Esposito, {\tscors Canonical and Perturbative Quantum
Gravity}, Preprint SISSA 10/93/A, (International School for Advanced
Studies, Trieste), Jan. 1993.

\endref
\bye